\documentclass[
reprint, 
 superscriptaddress,
 longbibliography,
 aps,
 prb,
]{revtex4-2}
\usepackage{caption}
\usepackage{graphicx}
\usepackage{dcolumn}
\usepackage{amsmath}
\usepackage{xspace}
\usepackage[T1]{fontenc}       
\usepackage{amssymb}
\usepackage{bm}
\usepackage[breaklinks=true,colorlinks,citecolor=blue,linkcolor=blue,urlcolor=blue]{hyperref}

\usepackage{tikz,xcolor,hyperref}
\usepackage[utf8]{inputenc}
\usepackage{makecell}
\usepackage[T1]{fontenc}
\usepackage{etoolbox}
\usepackage{txfonts}
\usepackage{xcolor}
\usepackage{fixltx2e}
\usepackage{ragged2e}
\usepackage{soul}

\makeatletter
\def\@email#1#2{%
 \endgroup
 \patchcmd{\titleblock@produce}
  {\frontmatter@RRAPformat}
  {\frontmatter@RRAPformat{\produce@RRAP{*#1\href{mailto:#2}{#2}}}\frontmatter@RRAPformat}
  {}{}
}%
\makeatletter

\renewcommand\section{\@startsection{section}{1}{\z@}%
  {0.8cm \@plus1ex \@minus .2ex}%
  {0.4cm}%
  {\normalfont\bfseries\raggedright}}

\renewcommand\subsection{\@startsection{subsection}{2}{\z@}%
  {0.6cm \@plus1ex \@minus .2ex}%
  {0.3cm}%
  {\normalfont\normalsize\bfseries\raggedright}}

\renewcommand\subsubsection{\@startsection{subsubsection}{3}{\z@}%
  {0.5cm \@plus1ex \@minus .2ex}%
  {0.2cm}%
  {\normalfont\normalsize\itshape\raggedright}}

\makeatother

\begin{document}

\preprint{AIP/123-QED}

\title{Phonon-Assisted Photoluminescence and Ultrafast Exciton Dynamics in Two-Dimensional Silicon Carbide}
\author{Afreen Anamul Haque}
\affiliation{Department of Electronics \& Electrical Communication Engineering, Indian Institute of Technology Kharagpur, West Bengal 721302, India}
\author{Rishabh Saraswat}
\affiliation{Nanoscale Electro-thermal Laboratory, Department of Electronics and Communication Engineering, Indian Institute of Information Technology Allahabad, Uttar Pradesh 211015, India}
\author{Aniket Singha}
\affiliation{Department of Electronics \& Electrical Communication Engineering, Indian Institute of Technology Kharagpur, West Bengal 721302, India}
\author{Rekha Verma}
\affiliation{Nanoscale Electro-thermal Laboratory, Department of Electronics and Communication Engineering, Indian Institute of Information Technology Allahabad, Uttar Pradesh 211015, India}
\author{Sitangshu Bhattacharya}
\altaffiliation{All correspondence should be addressed to Dr. Sitangshu Bhattacharya, \href{}{sitangshu@iiita.ac.in}}
\affiliation{Electronic Structure Theory Group, Department of Electronics and Communication Engineering, Indian Institute of Information Technology Allahabad, Uttar Pradesh 211015, India}



\begin{abstract}
\textbf{Abstract:}
\noindent Phonon-assisted photoluminescence provides a direct window into exciton–phonon interactions in semiconductors. Using fully ab initio many-body perturbation theory, including finite-momentum Bethe–Salpeter calculations, we investigate phonon-assisted emission and exciton dynamics in two-dimensional (2D) hexagonal silicon carbide (h-SiC) and benchmark its response against 2D h-boron nitride (h-BN). By explicitly resolving exciton–phonon matrix elements, we identify an electron-phonon scattering channel mediated by A$^\prime$ high-energy longitudinal and transverse optical phonons as the dominant contributors to sideband formation and quantify their spectral weights. We find that h-SiC exhibits pronounced phonon-assisted sidebands comparable to h-BN, despite a smaller exciton–phonon energy separation and fewer resolved replicas. The bright \textbf{K}–\textbf{K} exciton governs near-UV zero-phonon emission, while intervalley excitons acquire radiative character through symmetry-allowed optical-phonon coupling. Temperature-dependent scattering rates reveal an ultrashort bright-exciton lifetime of approximately 300 fs at 10 K, highlighting rapid exciton relaxation driven by intrinsic phonon channels.
\end{abstract}
\maketitle
\section{\label{sec:level1}INTRODUCTION}
\vspace{-0.3cm}
\noindent In semiconductors, the fundamental optical excitations are not free carriers but tightly bound excitons: electron–hole pairs that govern light–matter interactions at the microscopic level \cite{knox1963, Kira2006}. Their signatures are particularly pronounced in reduced-dimensional systems, notably two-dimensional (2D) materials \cite{poem2010, he2014, dey2016, wang2018colloquium}. Compared to their bulk counterparts, vanishing dielectric screening enhances the Coulomb interaction between the pair, giving rise to excitons with exceptionally large binding energies and long lifetimes. Consequently, both radiative and non-radiative optical processes in such systems are therefore intrinsically exciton-driven. Beyond shaping emission spectra, exciton–phonon (\textit{exc–ph}) interactions also regulate relaxation pathways and recombination kinetics, thereby determining the timescale of ultrafast exciton dynamics. These properties collectively constitute a central metric for the design of high-performance optoelectronic light emitters \cite{koperski2015, aharonovich2016, Marco2018, gu2019, sortino2021}.\\
\noindent Over the past two decades, photoluminescence (PL) experiments on ultra-clean materials including bulk hexagonal boron nitride (h-BN) \cite{Cassabois2016, shima2024} and 2D transition-metal dichalcogenides (TMDCs) \cite{Splendiani2010, mak2012, jones2013, zeng2013, Tonndorf2013, Yan2014, li2014, wang2015, chow2017, mueller2018, robert2020, funk2021, Roy2024} have revealed remarkably strong excitonic recombination channels accompanied by rich many-body optical phenomena. Irrespective of whether the quasiparticle band gap is direct or indirect, it is ultimately the optical gap within the excitonic dispersion that determines the accessibility of radiative and non-radiative emission pathways. In semiconductors characterized by an indirect optical gap, direct excitonic recombination is suppressed by momentum and/or spin selection rules. Several tungsten-based 2D TMDCs in their 1H phase \cite{Humberto2013, Palummo2015, robert2016, Molas2017, robert2017, Malic2018, thomas2021, Pablo2022, chen2022}, for example, possess a direct quasiparticle gap at the \textbf{K} valleys; however, their lowest-energy excitons are spin-forbidden (dark) within the excitonic dispersion, rendering the optical gap effectively indirect. Such indirect emission channels are typically bottlenecked by exciton thermalization and scattering processes at elevated temperatures, and therefore dominate at low temperatures \cite{Brem2020}. In contrast, molybdenum-based TMDCs exhibit both a direct quasiparticle and direct optical gap at \textbf{K} \cite{Splendiani2010, Zhang_2014, huang2023}.\\ 
\noindent This general paradigm breaks down in bulk h-BN \cite{Cassabois2016, Schue2016, shima2024}. Despite an indirect quasiparticle band gap and an indirect lowest-energy exciton, h-BN displays extraordinarily intense ultraviolet PL \cite{shima2024}, dominated by well-resolved phonon replicas at approximately 5.76 and 5.86 eV \cite{Cassabois2016}. This anomalously strong indirect emission surpassing even that of diamond \cite{Schue2016} originates from exceptionally robust \textit{exc–ph} coupling. State-of-the-art \textit{ab initio} investigations have firmly established the microscopic origin of this behavior through the solution of the finite-momentum (\textbf{Q}) Bethe–Salpeter Equation (BSE), explicitly capturing the phonon sidebands responsible for phonon-assisted radiative recombination \cite{Sponza2018, paleari2019exciton, shen2020, Chen2020}. Importantly, these same \textit{exc-ph} interactions that generate phonon replicas also dictate exciton relaxation and lifetime broadening within the excitonic manifold.\\
\noindent Here, we demonstrate that efficient phonon-assisted emission is not unique to h-BN. We show that 2D hexagonal silicon carbide (h-SiC) exhibits well-resolved phonon replicas extending from the UV-A into the visible spectral range, with emission intensities comparable to those reported for 2D h-BN. This indicates that the strong phonon-assisted emission characteristic of hBN is not strictly limited to wide-bandgap systems, but can also be effectively observed at the lower optical transition energies intrinsic to 2D h-SiC. Experimental realization of 2D h-SiC has recently been achieved via bottom-up synthesis on lattice-matched carbide substrates, most notably TaC(111) \cite{Polley2023}. First-principles local-density approximation calculations \cite{sun2008}, corroborated by angle-resolved photoemission spectroscopy (ARPES) \cite{Audunn2024}, confirm a planar honeycomb lattice with a direct quasiparticle band gap of approximately 3.90–4.19 eV at the \textbf{K} valley \cite{Sahin2009, bekaroglu2010}. These results place h-SiC on solid experimental and theoretical footing for investigating both phonon-assisted emission and ultrafast exciton relaxation dynamics. Furthermore, intrinsically large Born effective charges and pronounced lattice anharmonicity \cite{guo2018}, together with high chemical and mechanical stability \cite{bekaroglu2010}, suggest that 2D h-SiC is an attractive candidate for investigating phonon-assisted light-emission mechanisms.\\
\noindent We present a comprehensive \textit{ab initio} analysis of \textit{exc–ph} coupling in 2D h-SiC, establishing this interaction as a central mechanism governing indirect emission, exciton scattering, and lifetime renormalization. The exciton dispersion is analyzed in detail, and the interplay between excitonic effects and phonon-mediated processes including phonon-assisted radiative recombination and temperature-dependent exciton lifetimes is elucidated from first principles and group-theoretic arguments. The Supplementary Information (SI) \cite{Supplemental} contains all descriptive details, definitions and related convergence tests.
\noindent 
\begin{figure*}[!ht]
  \centering
  \includegraphics[width=1.00\textwidth]{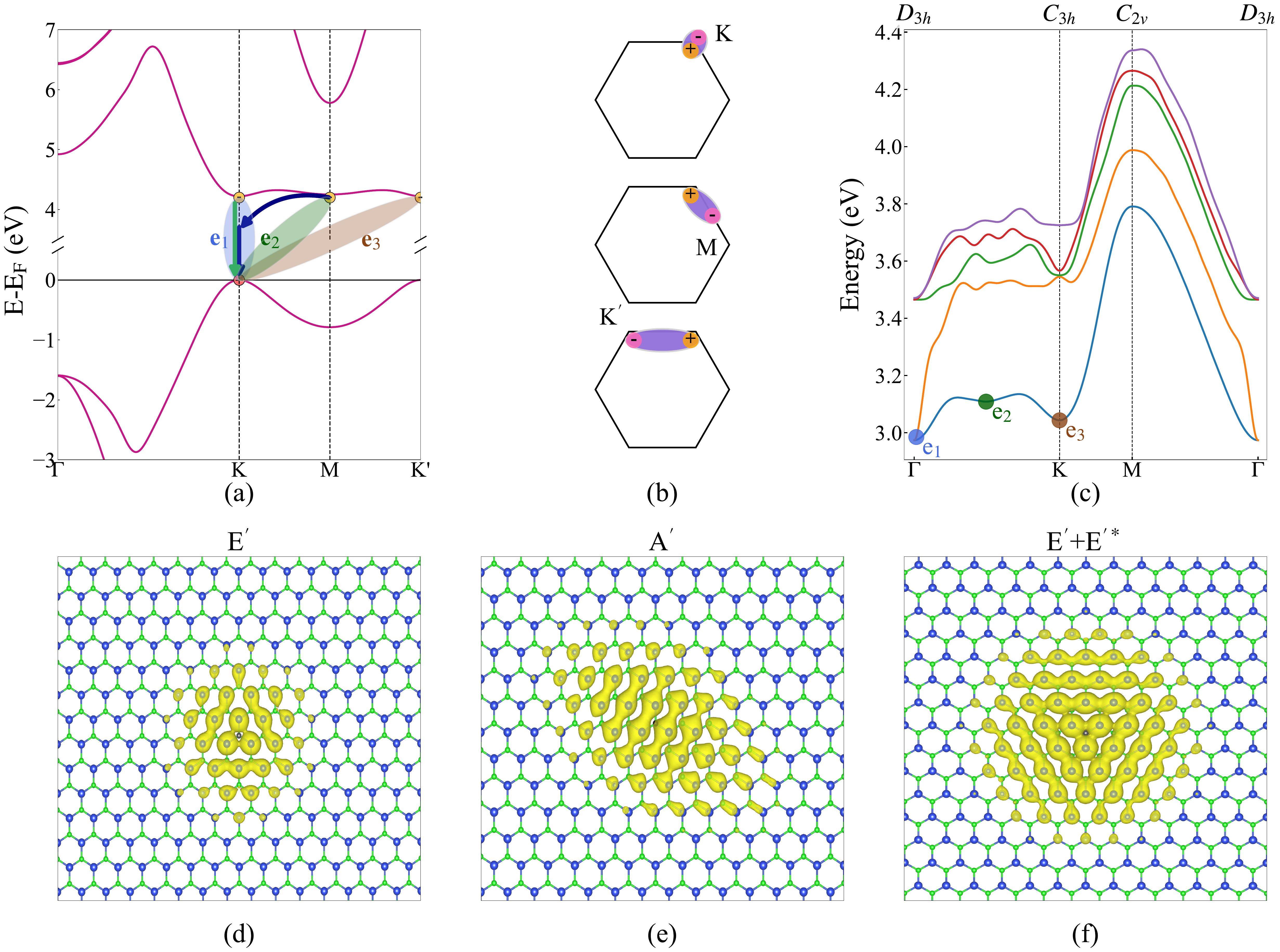}
  \caption{\justifying
Finite-\textbf{Q} exciton recombination in 2D h-SiC. (a) Quasiparticle band structure of 2D h-SiC computed within the G$_0$W$_0$ approximation. The dominant interband transitions forming the lowest excitons are indicated: $e_1$ corresponds to the direct \textbf{K}–\textbf{K} transition, while $e_2$ and $e_3$ denote indirect \textbf{M}–\textbf{K} and \textbf{K}–\textbf{K$^{\prime}$} transitions, respectively. The fancy-arrow schematically represent indirect recombination dynamics. 
(b) Momentum-space schematic of excitonic configurations in the first BZ, contrasting direct intravalley and intervalley (finite \(\mathbf{Q}\)) excitons. The corresponding real-space electron–hole probability densities are derived from the excitonic wavefunction. 
(c) Excitonic dispersion obtained from the solution of BSE, showing the five lowest branches along high-symmetry directions. The bright intravalley \textbf{K}–\textbf{K} exciton ($e_1$) occurs at \(\mathbf{Q}=0\) ($\boldsymbol{\Gamma}$), while finite-\(\mathbf{Q}\) states near \textbf{M} correspond to momentum-indirect excitons requiring phonon assistance for radiative recombination. 
(d–f) Real-space electron probability densities for the three lowest excitons ($e_1$, $e_2$, $e_3$), with the hole fixed on top of the carbon atom. The direct exciton ($e_1$) is compact and nearly isotropic, whereas the intervalley exciton ($e_3$) exhibits extended and anisotropic character consistent with their indirect nature.}
  \label{fgr:figure1}
\end{figure*}

\vspace{-0.3cm}
\begin{figure}[!ht]
  \centering
  \includegraphics[width=1\columnwidth, keepaspectratio]{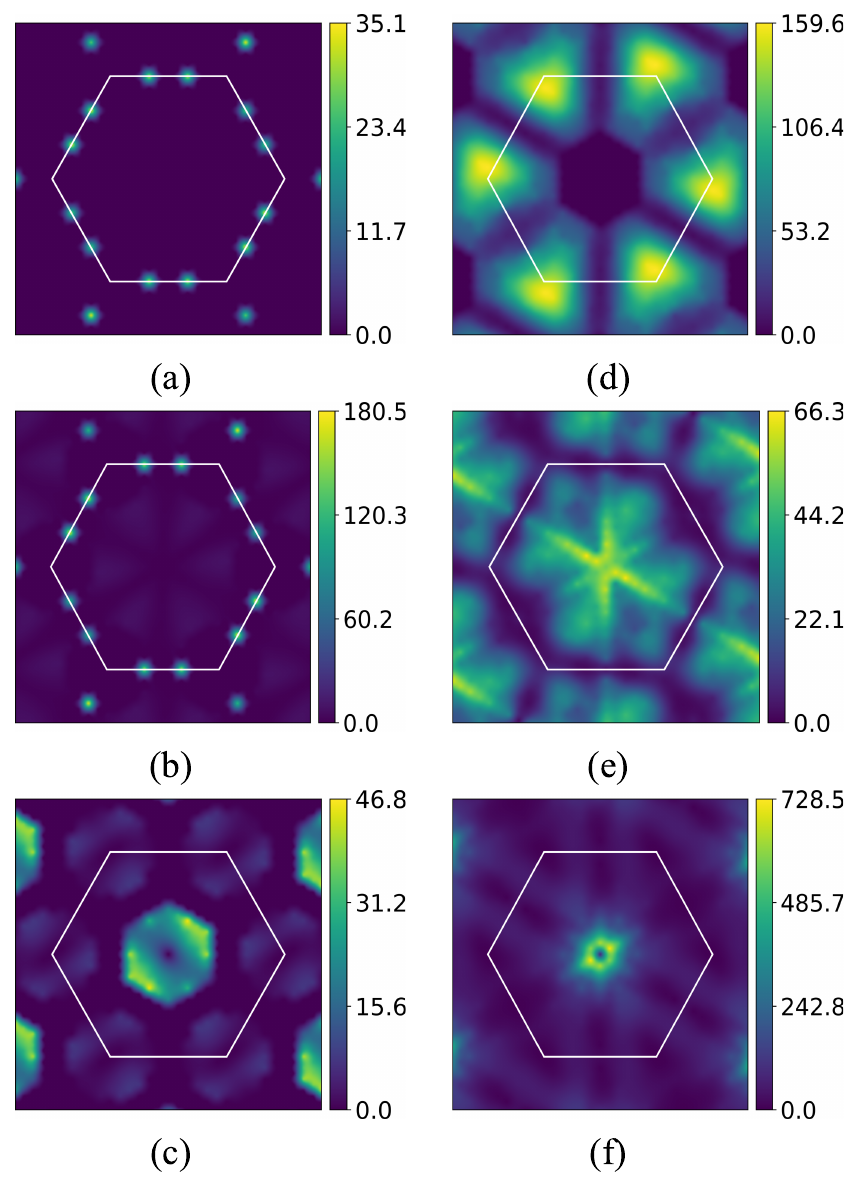}
  \caption{\justifying Phonon mode resolved \textit{exc}-\textit{ph} couplings in 2D h-SiC. (a)-(f) Phonon-mode-resolved exciton-phonon coupling strength \( \left|G_{\beta\lambda,\nu}(\mathbf{Q},\mathbf{q})\right|^2 \) in meV$^2$ mapped across the 2D BZ of the phonon wavevector $\mathbf{q}$. For these plots, the initial exciton is fixed as the fundamental bright state ($\beta = e_1$, $\mathbf{Q}_{in} = \mathbf{\Gamma}$), and the coupling is summed over all available final exciton states ($\lambda$). The white hexagon denotes the first BZ boundary. To preserve the relative physical magnitudes between distinct phonon branches, all absolute coupling values are uniformly scaled by a factor of 1e4 rather than being independently normalized. Consequently, the varying limits on the color bars reflect the true differences in absolute scattering probabilities mediated by each mode.}
  \label{fgr:figure2}
\end{figure}
\vspace{-0.3cm}

\section{RESULTS}{\label{sec_2}}
\vspace{-0.1cm}
\subsection{Excitonic symmetry and texture}{\label{sec_2a}}
\vspace{-0.1cm}
\noindent Within DFT, 2D h-SiC is identified as a polar wide–bandgap semiconductor with a bare indirect band gap of approximately 2.49 eV. Incorporating many-body effects at the single-shot GW (G$_0$W$_0$) level substantially renormalizes the spectrum, yielding a quasiparticle direct gap of 4.22 eV at the high-symmetry \textbf{K} valley of the BZ, as shown in Fig.~\ref{fgr:figure1}(a). The indirect character of the fundamental gap arises from the separation in momentum space between the valence-band maximum and conduction-band minimum. In contrast, the direct transition at \textbf{K} defines the lowest vertical excitation energy at the quasiparticle level and therefore establishes the fundamental energy scale for exciton formation and radiative optical processes.

\noindent The optical response is further constrained by symmetry. 2D h-SiC belongs to the $D_{3h}$ ($\bar{6}2m$) point group, within which the dipole operator transforms as $E^{\prime}\{x,y\}\oplus A_2^{\prime\prime}\{z\}$. In the present study, we focus on in-plane polarized light, which transforms according to the two-dimensional $E^{\prime}$ representation. For an optical transition from the totally symmetric ground state to be allowed, the final excitonic state must share the symmetry of the interacting light. Consequently, the optically active excitons at the $\mathbf{\Gamma}$ valley must transform according to the $E^{\prime}$ irreducible representation (irrep) and are thus doubly degenerate. To access momentum-indirect excitonic channels, we solve the finite-\textbf{Q} BSE, explicitly resolving the five lowest excitonic states across the BZ. The resulting dispersion, displayed in Fig.~\ref{fgr:figure1}(c), demonstrates that the bright $\mathbf{\Gamma}$ exciton ($e_1$) is the lowest exciton at $\mathbf{Q}=0$. This exciton is primarily characterized by transitions between the carbon atomic $p_z$ orbital at the valence band maximum and the silicon atomic $p_z$ orbital at the conduction band minimum (Fig.~S8 of SI \cite{Supplemental}). Because it is formed by two states that are odd under horizontal reflection ($\sigma_h$), the resulting exciton is even, correctly placing it in the two-dimensional $E^{\prime}$ irrep of $D_{3h}$, in perfect agreement with its optical activity for in-plane polarization. Away from $\mathbf{\Gamma}$, the degeneracy is lifted and the finite-\textbf{Q} excitons become nondegenerate. In particular, the states labeled $e_2$ and $e_3$ originate from intervalley electron–hole configurations involving \textbf{M}–\textbf{K} and \textbf{K}–$\textbf{K}^{\prime}$ momentum transfer, respectively. Their microscopic character is mapped in Fig.~\ref{fgr:figure1}(a), where the relevant interband transitions are indicated on the G$_0$W$_0$ band structure, and in Fig.~\ref{fgr:figure1}(b), which shows the corresponding electron–hole distributions in reciprocal space.\\
\noindent The symmetry properties of the finite-\textbf{Q} excitons are determined using compatibility relations that connect the irrepss at $\mathbf{\Gamma}$ to those of the little groups along high-symmetry directions \cite{Dresselhaus2008}. Along the $\mathbf{\Gamma}$–\textbf{K} direction, where the little group reduces to $C_s(m)$, the representations split into one-dimensional irreducible component ($A'$), lifting the degeneracy. At $\mathbf{Q}=\textbf{K}$, where the little group is $C_{3h}$, the excitonic states transform according to one-dimensional complex conjugate representations ($E^{\prime}$ and $E^{\prime*}$), consistent with the absence of symmetry-protected degeneracy. The corresponding symmetry classification is presented in Tables S2-S4 of the SI \cite{Supplemental}.

\noindent 2D h-BN and h-SiC display fundamentally distinct excitonic landscapes, reflecting pronounced differences in quasiparticle gap magnitude, dielectric screening strength, and the resulting electron–hole correlations. In the former, the first optically active exciton emerges at approximately 6.10 eV from interband transitions near the \textbf{K} point, placing it deep within the ultraviolet (DUV) regime \cite{Wirtz2006, Galvani2016}. In the latter, the lowest-energy bright exciton appears at approximately 2.97 eV, situated in the near-UV spectral window. This substantial spectral separation is primarily governed by the significantly larger quasiparticle gap of h-BN ($\sim$6.86 eV) relative to SiC ($\sim$4.22 eV), which shifts the entire excitonic manifold of the former to considerably higher energies.

\noindent Beyond this energetic offset, the nature of the excitons differs qualitatively. The weak in-plane dielectric screening and large band gap of the former promote strongly localized, Frenkel-like excitons with binding energies of 2.71 eV \cite{Wirtz2006, Galvani2016, Arnaud2006}. In contrast, the comparatively smaller gap and enhanced screening in the latter give rise to more spatially extended, Frenkel-like excitons \cite{YuCardona2010, Kira2006} with lower binding energies of approximately 1.17 eV. These differences are not merely quantitative but fundamentally reshape the excitonic phase space available for radiative recombination and phonon-assisted processes. Accordingly, the former is intrinsically suited to deep-UV excitonic emission, whereas the latter naturally operates in the near-UV regime, where exciton dispersion, binding strength, and coupling to lattice vibrations follow a distinctly different hierarchy.

\vspace{-0.3cm}
\subsection{Exciton-phonon couplings and scattering rates}
\begin{figure}[!ht]
  \centering
  \includegraphics[width=1\columnwidth]{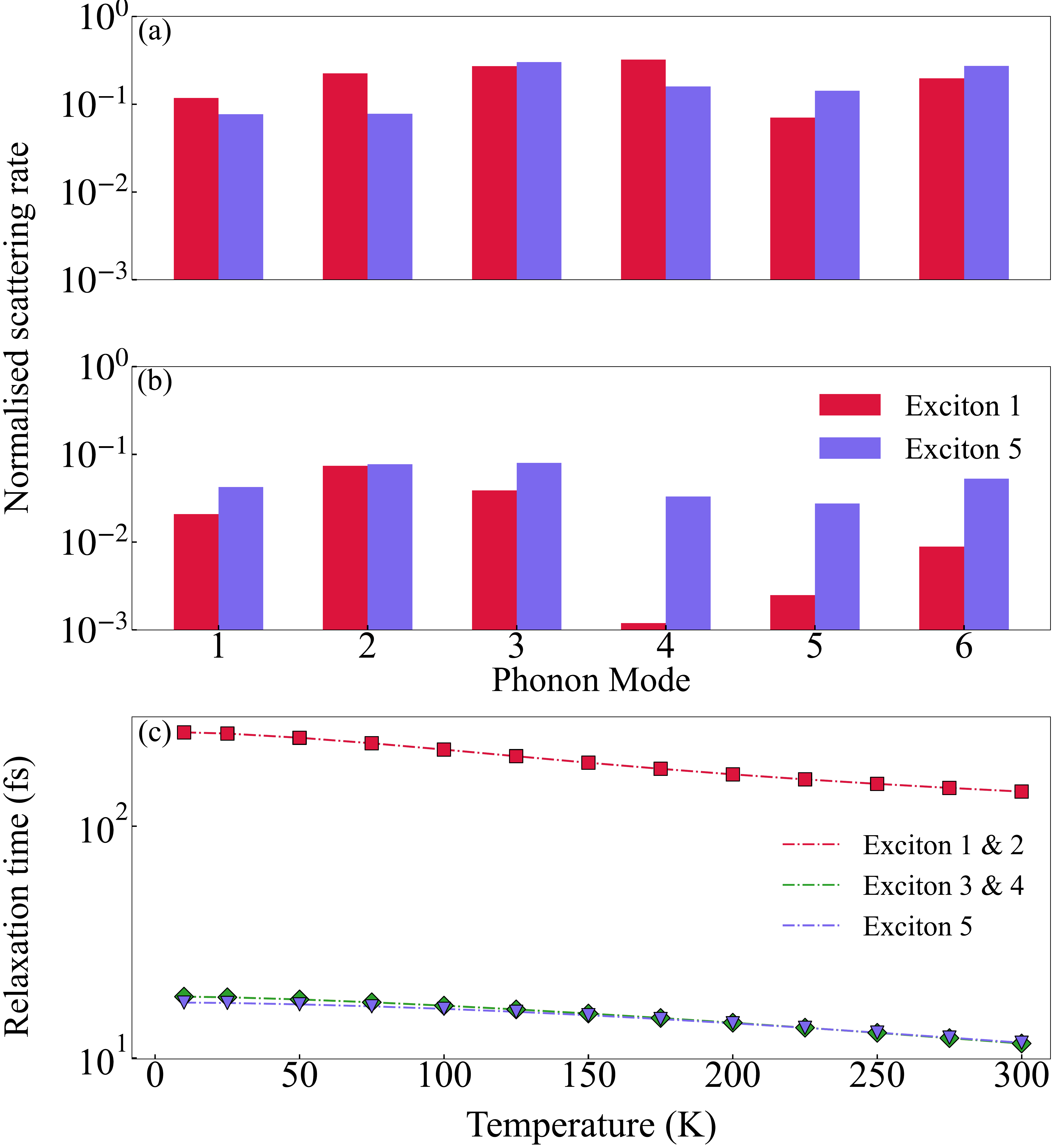}
  \caption{\justifying Phonon mode resolved scattering rates and relaxation time in 2D h-SiC. Phonon-mode-resolved normalized scattering rates (in log) of excitons~1 and~5 at (a) 10~K and (b) 300~K, respectively. (c) Relaxation time (in log) for the lowest five excitons (1–5) at $\mathbf{Q}=0$ ($\mathbf{\Gamma}$) as a function of excitonic temperature $T_{\mathrm{exc}}$. Additionally, the $e_1$ valley exciton introduced in Fig. \ref{fgr:figure2}(a) physically corresponds to the primary bright state among the lowest $\mathbf{\Gamma}$-point eigenstates (1–5) analyzed here and in Fig. \ref{fgr:figure4}.}
  \label{fgr:figure3}
\end{figure}

\begin{figure}[!ht]
  \centering
  \includegraphics[width=1\columnwidth]{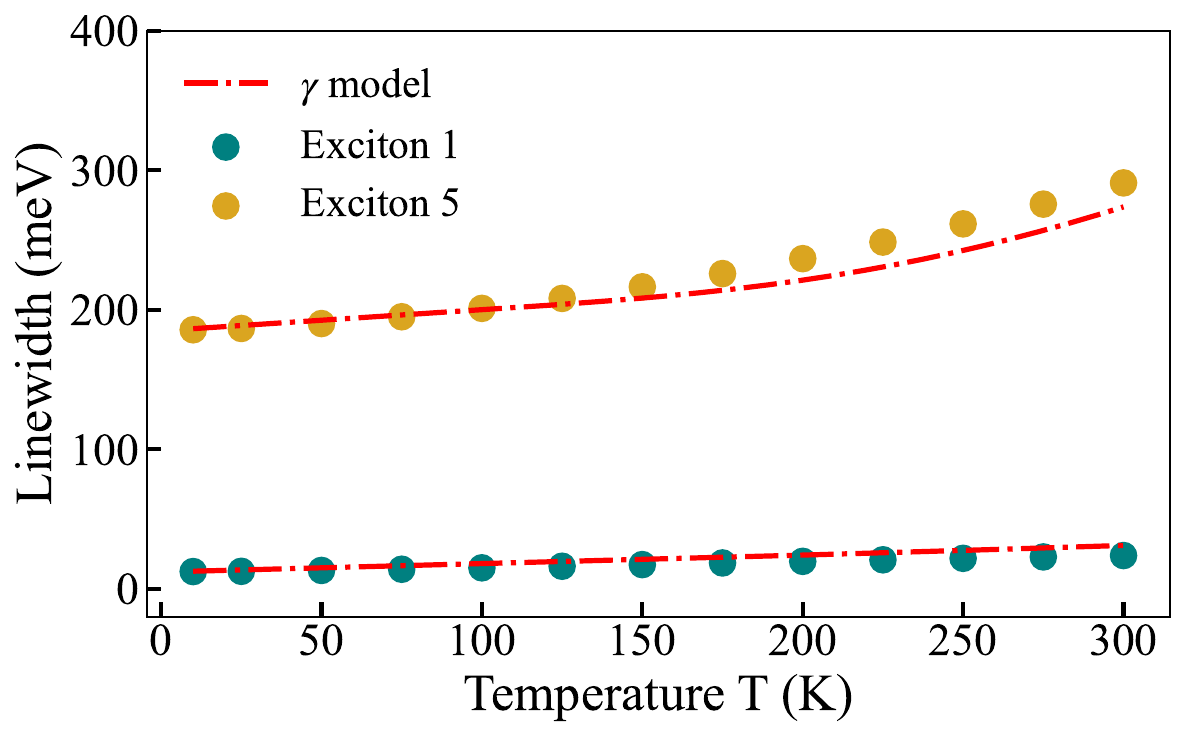}
  \caption{\justifying
Excitonic linewidth variation with temperature. Excitonic temperature (T = T$_{\mathrm{exc}}$) dependence of the linewidths of excitons~1 and~5 at the $\mathbf{\Gamma}$ valley. The red dashed line represents the $\mathbf{\gamma}$ \textit{ab initio} model.}
  \label{fgr:figure4}
\end{figure}

{\label{sec_2b}}
\vspace{-0.1cm}
\noindent To quantify \textit{exc–ph} interaction strengths and their impact on exciton relaxation dynamics, we first evaluate the \textit{el-ph} coupling matrix elements \cite{Giustino2017, Lechifflart2023}
\begin{equation}
g_{mn,\nu}(\mathbf{k},\mathbf{q})=\langle \mathbf{mk} \mid \Delta V_{\nu\mathbf{q}} \mid n(\mathbf{k}-\mathbf{q})\rangle,
\label{eqn:eqn1}
\end{equation}
\noindent where $\left|g_{mn,\nu}(\mathbf{k},\mathbf{q})\right|^2$ represents the coupling strength for a phonon-mediated electronic transition. The perturbation operator $\Delta V_{\nu \mathbf{q}}$ corresponds to the phonon-induced modulation of the Kohn–Sham potential for branch $\nu$. 
Within the excitonic basis, the corresponding \textit{exc–ph} coupling is given by \cite{Fulvio2019,Lechifflart2023, Marini2024}
\begin{multline}
\mathcal{G}_{\beta\lambda,\nu}\!\left(\mathbf{Q},\mathbf{q}\right)
=\sum_{\upsilon,\upsilon',c,c',\mathbf{k}}
A_{\lambda,\mathbf{Q}}^{\upsilon,c,\mathbf{k}}
\!\left[g_{\upsilon\upsilon',\nu}\!\left(\mathbf{k}-\mathbf{Q},\mathbf{q}\right)\delta_{c,c'}\right]
A_{\beta,\mathbf{Q+q}}^{\upsilon',c',\mathbf{k}^{\ast}}\\
-\sum_{\upsilon,\upsilon',c,c',\mathbf{k}}
A_{\lambda,\mathbf{Q}}^{\upsilon,c,\mathbf{k}}
\!\left[g_{c'c,\nu}^{\ast}\!\left(\mathbf{k}+\mathbf{q},\mathbf{q}\right)\delta_{\upsilon,\upsilon'}\right]
A_{\beta,\mathbf{Q+q}}^{\upsilon',c',\mathbf{k+\mathbf{q}}^{\ast}}
\label{eqn:eqn2}
\end{multline}
\noindent where $\left|\mathcal{G}_{\beta\lambda,\nu}(\mathbf{Q},\mathbf{q})\right|^{2}$ determines the probability of phonon-mediated transitions between excitonic states. Physically, this interaction represents a coherent superposition of electron–hole scattering events weighted by the excitonic wave function in the transition basis \cite{Chen2020}.\\
\noindent To identify the microscopic origin of ultrafast relaxation channels, we map the \textit{exc–ph} coupling strengths for phonon modes~1–6 throughout the BZ, as shown in Fig.~\ref{fgr:figure2}(a–f), averaging over the five lowest excitonic states, corresponding to $(\beta,\lambda)=(1,1),(1,2),(1,3),(1,4),(1,5)$. Optical phonon mode~6 (LO) exhibits maximum coupling intensity near the $\mathbf{\Gamma}$ valley, while mode~5 (TO) shows enhanced interaction along the $\mathbf{\Gamma}$–\textbf{M} direction, indicating selective momentum-dependent scattering pathways. \\
The corresponding exciton scattering rates are obtained from the imaginary part of the excitonic self-energy \cite{Lechifflart2023}:
\begin{multline}
\Xi_{\lambda}\!\left(\mathbf{Q}=0;\omega\right)
=\frac{1}{\Omega_{\mathrm{BZ}}}\sum_{\mathbf{q},\nu,\beta}
\mathcal{G}_{\beta\lambda,\nu}\!\left(\mathbf{Q}=0,\mathbf{q}\right)
\mathcal{G}_{\beta\lambda,\nu}^{\ast}\!\left(\mathbf{Q}=0,\mathbf{q}\right)\\
\times\!\left[
\frac{1+n_{\mathbf{q},\nu}}{\omega-E_{\mathbf{q},\beta}+\omega_{\mathbf{q},\nu}+i\eta}
+\frac{n_{\mathbf{q},\nu}}{\omega-E_{\mathbf{q},\beta}-\omega_{\mathbf{q},\nu}+i\eta}
\right]
\label{self_energy}
\end{multline}
\noindent where $\Omega_{\mathrm{BZ}}$ is the excitonic BZ volume,  $n_{\mathbf{q},\nu}$ is the Bose–Einstein occupation factor and excitonic temperature $T_{\mathrm{exc}}$ enters through the Boltzmann factor \cite{Fulvio2019}. The imaginary part of Eq.~(\ref{self_energy}) directly yields the phonon-limited exciton lifetime, thereby linking \textit{exc-ph} coupling to ultrafast relaxation dynamics. Only the diagonal self-energy contributions are retained in evaluating the total scattering rates.\\
\noindent As shown in Fig.~\ref{fgr:figure3}(a), the bright exciton ($n=1$), located near $\boldsymbol{\Gamma}$, exhibits comparatively strong \textit{exc–ph} scattering at low temperature, resulting in a phonon-limited relaxation time $\sim$ 260~fs in Fig. \ref{fgr:figure3}(c). This timescale defines an intrinsically ultrafast excitonic response governed by symmetry-allowed phonon channels.
With increasing temperature, enhanced phonon occupation strengthens scattering and reduces the relaxation time to below 150 fs at room temperature. Higher-lying dark excitons (n=3–5) display substantially shorter relaxation times (on the order of 15~fs or less) already at low temperature, reflecting stronger coupling to available phonon modes and more efficient redistribution within the excitonic manifold. The mode-resolved scattering rates (Figs.~\ref{fgr:figure3}(a and b)) clarify the microscopic origin of this behavior: at 10~K, all phonon branches contribute equally, whereas at 300~K, low-energy acoustic phonons provide the primary scattering channel for the lowest bright exciton. This stands in stark contrast to the temperature-dependent redistribution of excitonic populations reported in wide-bandgap 2D systems such as h-BN \cite{Lechifflart2023, Schue2016}, WS$_2$ \cite{ye2014, Brem2020}, WSe$_2$ \cite{Brem2020} and AlN \cite{Pushpendra2025} where phonon-assisted recombination from momentum-dark states becomes prominent at low temperatures due to a narrow exciton energy distribution and restricted access to bright states.
\\
\noindent In Fig.~\ref{fgr:figure4}, we report the non-radiative linewidths $\gamma$ of the first and fifth excitons. The temperature dependence of the non-radiative broadening can be phenomenologically described as \cite{wang2018colloquium}
\begin{equation}
\gamma_{\mathrm{NR}}(T) = \gamma_{0} + \gamma_{\mathrm{ac}}\,T
+ \frac{\gamma_{\mathrm{op}}}{\exp\!\left(\dfrac{\Omega}{k_{B}T}\right)-1},
\end{equation}
where $\gamma_{\mathrm{ac}}$ and $\gamma_{\mathrm{op}}$ quantify the exciton–acoustic and exciton–optical phonon coupling strengths, respectively, $\gamma_{0}$ denotes the residual dephasing rate at $T=0$~K, and $\Omega$ represents an effective optical phonon energy. The calculated broadening for exciton~1 exhibits an approximately linear temperature dependence across the full range considered, indicating that optical phonon contributions are negligible and that acoustic phonon scattering dominates from 10 to 300~K. For exciton~5, the broadening is linear at lower temperatures, whereas at higher temperatures, optical phonon contributions begin to dominate, resulting in an onset of exponential growth governed by the Bose distribution. Fitting the data yields $\gamma_{0}=12$~meV for exciton~1 and $\gamma_{0}=185$~meV for exciton~5, with corresponding acoustic coupling coefficients of $\gamma_{\mathrm{ac}}=0.06$~meV/K and $0.15$~meV/K, respectively. The optical coupling coefficient for exciton~5 is determined to be $\gamma_{\mathrm{op}}=2$~eV.
\noindent While linewidth broadening captures the cumulative effect of \textit{exc-ph} scattering on non-radiative dephasing and relaxation, a complementary microscopic perspective is obtained by explicitly evaluating phonon-assisted optical emission within the excitonic formalism.
\noindent At low temperatures, the phonon population is strongly suppressed. Nevertheless, exciton relaxation and recombination remain efficient through phonon emission processes, particularly for modes with large \textit{el}-\textit{ph} matrix elements and for momentum-dark excitons that accumulate near the dispersion minimum. This low-temperature redistribution of excitonic populations directly influences the phonon-assisted emission spectrum.
\subsection{Phonon-assisted PL emission and symmetry constraints}{\label{sec_2c}}
\vspace{-0.1cm}
\noindent To quantitatively describe these processes, phonon-mediated satellites are evaluated using a generalized form of the Roosbroeck–Shockley relation \cite{Pillai2026}.
\begin{multline}
I^{PL}(\omega;T) \propto \frac{1}{N_{\mathbf{q}}} \sum_{\mu, \beta, \mathbf{q}} \Biggl|\sum_{\lambda}T_{\lambda}^* \mathcal{D}^{*,\pm}_{\beta \lambda, \mathbf{q}\mu}\Biggr|^2 \\
\times N_{\beta \mathbf{q}}(T_{\mathrm{exc}})F^{\pm}_{\mu \mathbf{q}}(T)\delta(\omega - [E_{\mathbf{q}, \beta} \mp \omega_{\mu \mathbf{q}}])
\label{PL}
\end{multline}
\noindent where $T_{\lambda}$ denotes the dipole matrix elements of exciton~$\lambda$. The quantity $\left| \mathcal{D}^{\pm}_{\beta \lambda, \mathbf{q}\mu} \right|$ characterizes the phonon-assisted satellite strengths. $E_{\mathbf{q},\beta}$ $\lambda$ and $\beta$ mediated by phonon mode $(\mathbf{q},\nu)$, while $E_{\mathbf{q},\beta}$ denotes the energy of exciton~$\beta$ at finite momentum. $N_{\beta \mathbf{q}}$ denotes the Bose-Einstein occupation factor for these excitons. $F^{\pm}_{\mu \mathbf{q}}(T) = \frac{1}{2} \pm \frac{1}{2} + N_{\mu \mathbf{q}}(T)$, where $N_{\mu \mathbf{q}}(T)$ is the Bose-Einstein occupation factor for phonons.  The Boltzmann factors encode thermal population of excitonic states at an effective exciton temperature $T_{\mathrm{exc}}$.
\begin{figure}[!ht]
\centering
\includegraphics[width=1\columnwidth]{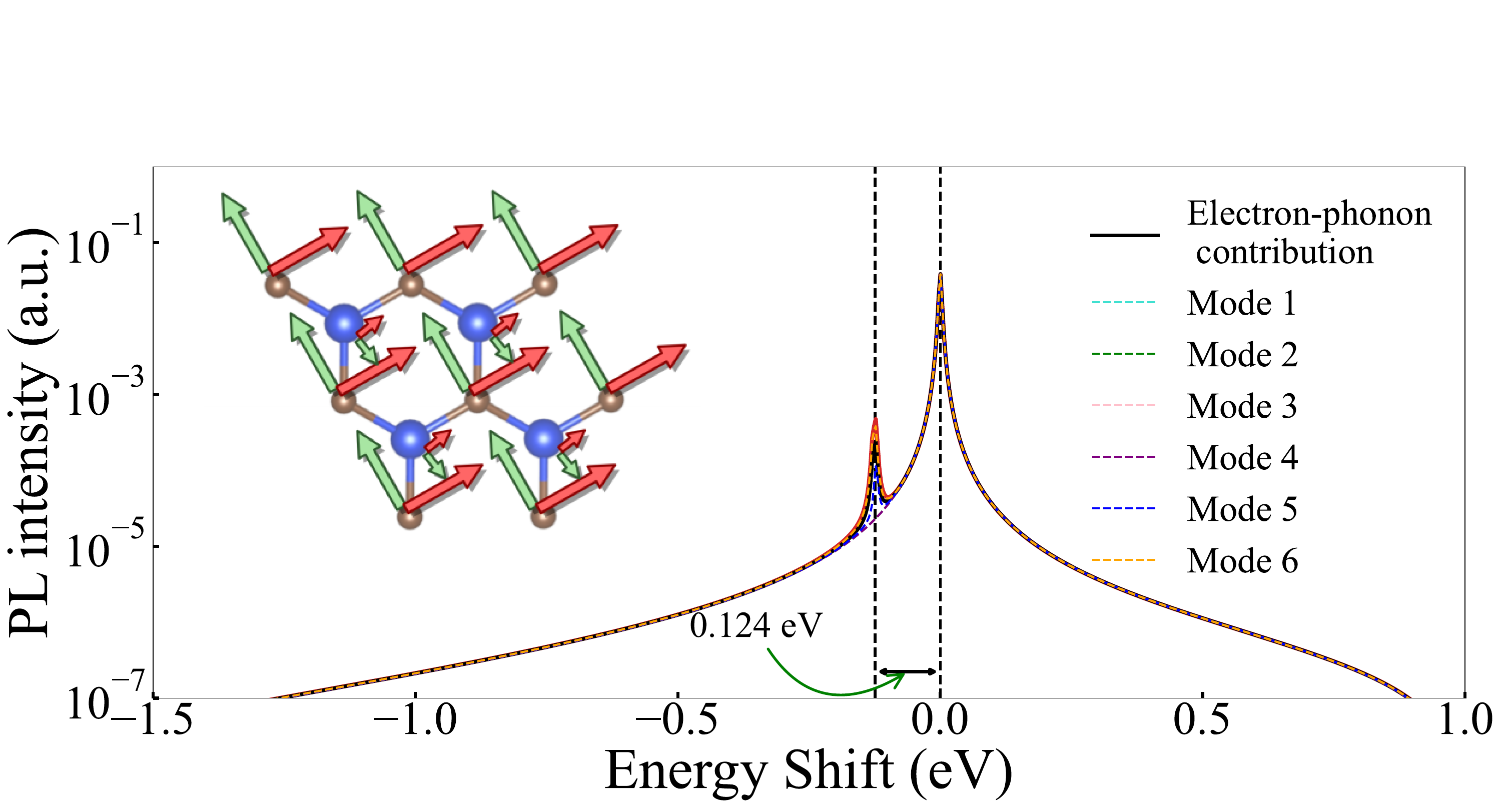}
\caption{\justifying 
Indirect PL emission in 2D h-SiC. Phonon-assisted PL plotted on a logarithmic scale. The $x$-axes has been shifted by the energy of the bright exciton (2.97~eV). PL spectra at an excitonic temperature of 10 K. The inset shows the atomic vibrations (Si atoms in blue, C atoms in brown) are  of mode 5(A$^{\prime}$) and 6(A$^{\prime}$) by green and red arrows respectively.} 
\label{fgr:figure5}
\end{figure}
\\ \noindent The bright exciton in h-SiC is found to constitutes the lowest-energy state, while higher-lying momentum-indirect excitons populate the dispersion. 
\begin{figure*}[!ht]
  \centering
  \includegraphics[width=1\textwidth]{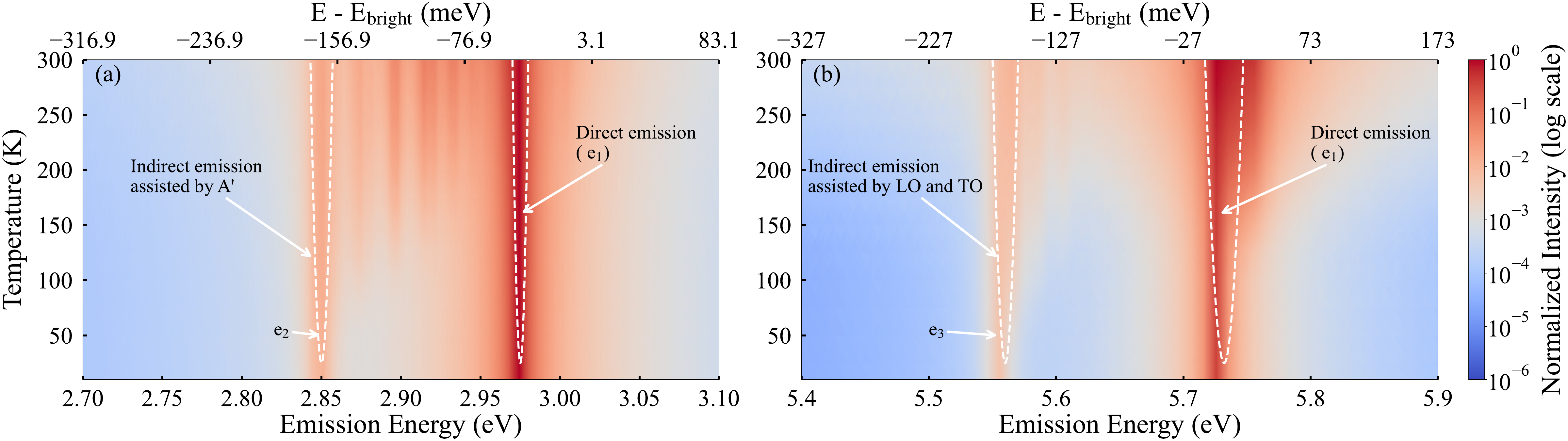}
  \caption {\justifying Phonon-assisted PL emission from 2D h-SiC and h-BN. Temperature-dependent normalized phonon-assisted PL spectra for (a) 2D h-SiC and (b) h-BN. The heatmaps display the PL intensity (logarithmic color scale) as a function of absolute emission energy (bottom axis) and detuning from the bright exciton $e_1$ (top axis) across a temperature range of $10$ to $300\,\mathrm{K}$. The spectra are normalized to their respective maximum emission peaks to emphasize the relative spectral weights of the excitonic and phonon-replica features. Dashed white lines trace the dominant emission channels, highlighting the direct emission ($e_1$) and the symmetry-allowed indirect emission ($e_2$), which is mediated by the $A^\prime$ phonon mode in 2D h-SiC and the LO-TO modes in h-BN. Moreover, the sideband peak in 2D h-SiC is $\sim$ 1.3 times larger than h-BN.}
  \label{fgr:figure6}
\end{figure*}
The energy offset between the direct and relevant indirect state amounts to 124~meV, setting the minimum phonon energy required to activate phonon-assisted recombination. From the calculated phonon spectrum, only high-energy optical modes fulfill this requirement, identifying $e_3$ as the most probable contributing exciton (see Fig.~S15 in the SI \cite{Supplemental}). In Fig. \ref{fgr:figure5} the resulting phonon-assisted PL spectrum is shown on a logarithmic scale. At 10~K, a distinct sideband emerges below the zero-phonon peak. Mode-resolved analysis demonstrates that this feature is governed exclusively by optical phonon modes~5 (TO) and~6 (LO) (blue and yellow dashed lines), with negligible contributions from the remaining branches. This selectivity arises not only from energetic compatibility but also from microscopic coupling: the in-plane bond stretching and compression of the TO-LO modes efficiently couple to the momentum-dark, in-layer-localized exciton along the $\mathbf{\Gamma}$–\textbf{K} direction, resulting in comparatively large \textit{exc–ph} matrix elements. In both panels of the figure, the energy axis is referenced to the bright-exciton energy (2.97~eV) to facilitate direct comparison of satellite positions. We assume a thermal equilibrium process, and the exciton populations are described by a Boltzmann distribution, consistent with the implementation employed in our calculations. Consequently, the phonon-assisted emission is much weaker ($\sim$ 100 times) than the zero-phonon emission. To clearly visualize these phonon-assisted features and facilitate comparison with previous theoretical studies on 2D h-BN, the photoluminescence spectra are presented on a logarithmic intensity scale (see Fig. S17 in the SI \cite{Supplemental}).\\
\noindent A comparative analysis between 2D h-SiC and h-BN is presented in Figs.~\ref{fgr:figure5}(a and b), respectively, where the direct and phonon-assisted emission features are highlighted by white dashed contours. Both the zero-phonon and satellite peaks of h-BN are blue-shifted relative to SiC, consistent with its larger quasiparticle band gap and higher excitonic energies. The heatmap further indicates that, even at low temperature, the phonon-assisted PL intensity of h-BN slightly exceeds that of SiC. With increasing temperature, SiC exhibits a noticeable enhancement in satellite population, whereas the indirect feature in h-BN remains comparatively sharper and more intense ( for example, see Fig. S16 and S17 of the SI \cite{Supplemental}.) 
\begin{table}[b]
\centering
\renewcommand{\arraystretch}{1.2}
\setlength{\tabcolsep}{8pt}
\begin{tabular}{|c|c|c|c|}
\hline
\textbf{Mode} & \textbf{Phonon irrep $\Gamma_{\mathrm{ph}}$} & \textbf{$\Gamma_{\mathrm{ph}}\otimes\Gamma_{\mathrm{light}}$} & \textbf{Transition} \\
\hline
1 & $A^{\prime\prime}$ & $A^{\prime\prime}$ & Forbidden\\
\hline
2 & $A^{\prime}$        & $A^{\prime}$ & Allowed\\
\hline
3 & $A^{\prime\prime}$        & $A^{\prime\prime}$ & Forbidden \\
\hline
4 & $A^{\prime}$        & $A^{\prime}$ & Allowed\\
\hline
5 & $A^{\prime}$ & $A^{\prime}$ & Allowed\\
\hline
6 & $A^{\prime}$        & $A^{\prime}$ & Allowed\\
\hline
\end{tabular}
\caption{\justifying Symmetry-allowed phonon-assisted optical transitions. Tensor-product decomposition between the in-plane light irrep, $\Gamma_{\mathrm{light}}=A^{\prime}$, and the phonon irreps at $\mathbf{K}$ for the six phonon branches considered. The last column represents the symmetry driven transitions.}
\label{tab:phonon}
\end{table}
\\ \noindent In bulk h-BN, the lowest-energy exciton is momentum-indirect and lies below the bright exciton, thereby governing the phonon-assisted emission, as established previously \cite{Cannuccia2019, Cassabois2016}. The corresponding PL spectrum resolves multiple phonon replicas associated with TA (64~meV), LA (95~meV), and dominant TO-LO modes (188~meV), referenced to the indirect-exciton energy. Although both materials exhibit TO-LO-mediated sidebands, their microscopic origins differ substantially. In 2D h-SiC, the satellites originate from a bright ground-state exciton, resulting in a comparatively smaller \textit{exc-ph} energy separation. In contrast, h-BN hosts replicas emerging from a lower-lying dark exciton, which leads to larger optical-phonon offsets and comparatively weaker acoustic contributions.
\noindent In contrast to monolayer MoS$_2$, where phonon-assisted recombination becomes symmetry-allowed only under biaxial strain that renders the lowest-energy optical transition momentum-indirect \cite{Saraswat2025}, 2D h-SiC supports such processes intrinsically. Its non-centrosymmetric lattice and pronounced polar character activate efficient \textit{exc-ph} coupling without the need for external symmetry breaking.\\
\noindent To formalize this symmetry activation, we perform a group-theoretical analysis (see SI \cite{Supplemental}), demonstrating that the states relevant to phonon-assisted PL in 2D h-SiC strictly transform according to the $A^\prime$ irrep of the $C_s(m)$ subgroup. In light of this, as shown in Table~\ref{tab:phonon}, symmetry considerations strictly forbid modes 1 and 3 from participating in the phonon-assisted PL. Furthermore, while modes 2 and 4 are symmetry-allowed, their low phonon energies are insufficient to account for the large energy separation of the observed macroscopic phonon sidebands. Therefore, only the high-energy optical modes 5 and 6 can contribute to the sidebands, in excellent agreement with the calculated emission spectra in Fig.~\ref{fgr:figure5}. The corresponding optical selection rules, derived from the tensor products of the light and phonon irreps, are summarized in Table~\ref{tab:phonon}.\\
\noindent Despite being symmetry-allowed, only a subset of modes contributes significantly to the observed emission. The calculated \textit{exc–ph} matrix elements indicate that, beyond the optical LO and TO branches, a few additional modes exhibit finite coupling along the $\mathbf{\Gamma}$–\textbf{K} direction. However, the spectral weight of the sidebands (Fig.~\ref{fgr:figure6}) is overwhelmingly dominated by the TO and LO modes. This dominance is dictated by energetic constraints: only phonons with sufficient energy to bridge the excitonic separation efficiently activate radiative channels, whereas acoustic modes lie well below this threshold and therefore carry negligible intensity. In bulk 4H–SiC, multiple phonon sidebands appear below the bright exciton due to free-exciton transitions, as reported by Lyu and Lambrecht \cite{Lyu2020} and Ivanov \textit{et al.} \cite{Ivanov1998}. In 2D h-SiC, the sideband structure instead reflects a strictly energy-selective \textit{exc-ph} matching condition.\\
\noindent In summary, we establish that phonon-assisted luminescence in 2D h-SiC emerges intrinsically from its polar, non-centrosymmetric lattice and underlying $D_{3h}$ symmetry, without reliance on external symmetry breaking or strain engineering. Finite-momentum BSE calculations combined with a fully \textit{ab initio} treatment of \textit{exc-ph} coupling demonstrate that while the bright \textbf{K}–\textbf{K} exciton governs zero-phonon emission near 2.97 eV, intervalley exciton (\textbf{K}–\textbf{M}) acquire radiative character through selective coupling to high-energy TO and LO phonon modes. The resulting sideband structure is dictated by strict energetic matching and symmetry-allowed scattering channels. Temperature-dependent scattering rates reveal a crossover from optical-phonon-dominated relaxation at cryogenic conditions to acoustic-phonon-mediated dynamics at elevated temperatures, providing a unified picture of exciton thermalization and recombination. In contrast to h-BN, h-SiC exhibits a similar strong \textit{exc-ph} coupling, phonon-sideband ratio, and fewer resolved replicas, consistent with the predominance of zero-phonon emission. These results identify 2D h-SiC as a symmetry-activated, finite-$\mathbf{Q}$ dark-exciton system governed by the $C_s(m)$ subgroup, where phonon-assisted recombination is inherently enabled. A viable route toward achieving highly efficient room-temperature emission in h-SiC could be the introduction of point defects, such as nitrogen-vacancy complexes or adjacent interstitials, which can generate deep localized states and function as quantum emitters. However, a rigorous theoretical treatment of such defect-assisted emission requires explicit calculations of charge-transition levels, configuration-coordinate diagrams, electron–phonon coupling, carrier capture coefficients, Huang--Rhys factor, and zero-phonon line \cite{Alkauskas2014}. These calculations remain as a future scope of the present work.
\section{Discussion}{\label{sec_3}}
\vspace{-0.3cm}
\noindent We have presented a comprehensive ab initio investigation of exciton–phonon interactions in two-dimensional hexagonal SiC, demonstrating that its strict $D_{3h}$ crystal symmetry and intrinsic inversion asymmetry fundamentally regulate phonon-assisted optical processes. The combination of finite-momentum Bethe–Salpeter analysis and mode-resolved exciton–phonon coupling establishes a coherent microscopic framework connecting symmetry, dispersion, and temperature-dependent relaxation dynamics. Beyond identifying specific scattering channels, this work clarifies how symmetry-allowed intervalley excitons acquire radiative character in a non-centrosymmetric two-dimensional semiconductor without external perturbations. The resulting emission landscape, characterized by limited but symmetry-selective phonon sidebands, distinguishes SiC from prototypical systems such as h-BN and TMDCs. More broadly, these findings provide theoretical insight into the role of intrinsic crystal symmetry in activating and constraining phonon-assisted recombination pathways in low-dimensional materials. We emphasize that the present study addresses the intrinsic excitonic and phonon-assisted optical properties of an ideal monolayer h-SiC. Effects arising from structural defects, substrate interactions, non-radiative recombination channels, and device-level quantum efficiency are beyond the scope of the present work and remain important directions for future experimental and theoretical investigations.
\section{\label{sec:level4} METHODS}
\vspace{-0.1cm}
\subsection{\label{sec:level4a} Ground state calculations}
\vspace{-0.1cm}
\noindent Ground-state electronic structure calculations are carried out within density functional theory (DFT) using the \textsc{Quantum ESPRESSO} package \cite{Giannozzi2017}, while excited-state properties are treated within many-body perturbation theory (MBPT) package \textsc{YAMBO} \cite{Sangalli2019}. Fully relativistic norm-conserving pseudopotentials \cite{hamann2013optimized} are employed throughout. Silicon and carbon atoms are described using [Ne] and [He] cores, with valence configurations 3s$^2$3p$^2$ and 2s$^2$2p$^2$, respectively. A vacuum spacing of approximately 20~\AA{} is introduced along the out-of-plane ($z$) direction to eliminate spurious interactions between periodic images. Ground-state convergence is achieved using a kinetic energy cutoff of 90~Ry and an 18 $\times$ 18 $\times$ 1 \textbf{k}-point mesh for Brillouin-zone (BZ) sampling. Structural relaxation yields an equilibrium in-plane lattice constant of 3.09~\AA{}, consistent with previously reported values for 2D h-SiC \cite{Polley2023, Sahin2009, bekaroglu2010}. All the relevant parameters have been reported in the Table S1 of the SI \cite{Supplemental}. 
\subsection{\label{sec:level4b} Electron-phonon calculations}
\vspace{-0.1cm}
\noindent Lattice dynamical properties are computed within density functional perturbation theory (DFPT) on a uniform 18$\times$18$\times$1 \textbf{q}-grid. The self-consistent energy threshold for phonon calculations is set to $10^{-17}$~Ry with a mixing factor of 0.7. Phonon dispersions are subsequently interpolated onto a dense 120 $\times$ 120 $\times$ 1 \textbf{q}-mesh to ensure high reciprocal-space resolution. Electron–phonon (\textit{el-ph}) self-energies are evaluated on commensurate \textbf{k} and \textbf{q} grids, including both Fan \cite{Fan1950} and Debye–Waller \cite{cannuccia2011effect} contributions. The \textit{el-ph} calculation results are shown in Figs. S1-S4 of the SI. \cite{Supplemental}.
\subsection{\label{sec:level4c} Excited state calculations}
\vspace{-0.1cm}
\noindent Quasiparticle corrections are computed within the single-shot GW approximation \cite{Schilfgaarde2006}. Corrections are applied to the first four valence and conduction bands, respectively. The exchange self-energy summation employs a cutoff of 50~Ry, while the irreducible polarization function is evaluated using 100 bands (4 occupied and 96 unoccupied). To remove long-range momenta divergences at all self-energy levels, BZ integrations are performed using a random integration method (RIM) \cite{Pulci1998, Rozzi2006}, sampling $10^6$ random momenta points. A slab Coulomb cutoff \cite{Guandalini2023} is applied along the out-of-plane direction to remove artificial screening, and dynamical screening is treated within the Godby–Needs plasmon-pole model \cite{Godby1989}. Convergence tests for the G$_0$W$_0$ parameters are reported in Figs.~S5-S8 of the SI \cite{Supplemental}.
\noindent The convergence results for the BSE solutions at $\mathbf{Q=0}$ are shown in Figs. S9-S12 and Fig. S18 in the SI \cite{Supplemental} respectively. 
\subsection{\label{sec:level4d} BSE with finite-\textbf{Q} calculations}
\vspace{-0.1cm}
\noindent We explicitly solve the BSE for excitons with finite center-of-mass momentum while keeping the lattice fixed at its equilibrium geometry (Fig. S19 in the SI \cite{Supplemental}). This finite-$\mathbf{Q}$ treatment is essential for a consistent description of momentum-indirect recombination pathways and the associated phonon sidebands, which are not accessible within a strictly $\mathbf{Q}=0$ formalism.  Exciton-phonon matrix elements are obtained by combining finite-momentum BSE solutions with DFPT phonon perturbations. We appropriately fix the gauge invariance appearing in the \textit{exc-ph} coupling strength (Eq. \ref{eqn:eqn2}) \cite{letzelphc, Yambopy}.
Five excitonic states at $\mathbf{Q}\neq 0$, indexed by $\beta$ in Eq.~(\ref{PL}), are selected as initial states. These states couple to phonons through \textit{exc–ph} scattering and relax into final excitonic states at $\mathbf{Q}=0$, indexed by $\lambda$ in Eq.~(\ref{PL}), consistent with the convergence analysis shown in Fig.~S13 of the SI \cite{Supplemental}. The \textit{exc-ph} coupling strength for the first 5 excitons is plotted in Fig. S14 of the SI \cite{Supplemental}. \\
The BSE kernel includes both direct (Hartree) and exchange terms in the electron-hole interaction, ensuring a consistent treatment of excitonic correlations. For numerical stability, an energy threshold of $0.1\,\mathrm{meV}$ is imposed on the Fan denominator, and a damping parameter of $2\,\mathrm{meV}$ is introduced in the evaluation of the \textit{exc–ph} self-energy. The PL emission spectra are computed using interpolated exciton and phonon energies on a $120\times120\times1$ grid.
\noindent Direct diagonalization of the full Hamiltonian to obtain the phonon-assisted PL spectrum is not feasible, since the formulation involves two distinct frequency-dependent components associated with electron-electron and electron-phonon interactions. Instead, phonon-assisted PL is derived within a Green's-function formalism through a first-order expansion of the finite-momentum exciton propagator, in which the dynamical \textit{exc-ph} kernel enters a Dyson-like equation. In practice, the DFPT matrix elements are subsequently used to construct the \textit{exc-ph} coupling terms and the associated self-energies within the MBPT framework \cite{Mahan2014}, while maintaining consistency between the transferred $\mathbf{Q}$ and phonon momenta $\mathbf{q}$ grids through additional convergence checks. Finally, the excitonic response is obtained by expanding the Dyson-like equation to first order. Consequently, the electronic band renormalization induced by phonons does not enter the finite-momentum excitonic BSE Hamiltonian.
\section{Declaration Statements:}
\section{Data Availability}
\vspace{-0.3 cm}
\noindent The data supporting the findings of this study are available within the paper and its Supplementary Information. The computational input and output files are available from the corresponding author upon reasonable request.
\section{Acknowledgements}
\vspace{-0.3 cm}
\noindent This work was carried out with financial support from the Anusandhan National Research Foundation (ANRF), India, under Grant No.~CRG/2023/000476. The authors acknowledge the National Supercomputing Mission facility \emph{PARAM Shivay} at IIT (BHU), India, for providing computational resources. Afreen Anamul Haque and Rishabh Saraswat acknowledge fellowship support from the Ministry of Education, Government of India.
\section{Author Contribution}
\vspace{-0.3 cm}
\noindent A.A.H., R.S., A.S., R.V., and S.B. contributed equally to this work. All authors have read and approved the final manuscript. Correspondence should be addressed to Dr. Sitangshu Bhattacharya, email id: \href{}{sitangshu@iiita.ac.in}
\section*{Competing Interests}
\vspace{-0.3 cm}
\noindent The authors declare no competing financial or non-financial interests.
\vspace{0.3 cm}

\bibliographystyle{naturemag}
\bibliography{References}

\clearpage
\onecolumngrid

\section*{Figure Legends}

\noindent \textbf{Figure 1.} Finite-\textbf{Q} exciton recombination in 2D h-SiC. (a) Quasiparticle band structure of 2D h-SiC computed within the G$_0$W$_0$ approximation. The dominant interband transitions forming the lowest excitons are indicated: $e_1$ corresponds to the direct \textbf{K}–\textbf{K} transition, while $e_2$ and $e_3$ denote indirect \textbf{M}–\textbf{K} and \textbf{K}–\textbf{K$^{\prime}$} transitions, respectively. The fancy-arrow schematically represent indirect recombination dynamics.
(b) Momentum-space schematic of excitonic configurations in the first BZ, contrasting direct intravalley and intervalley (finite \(\mathbf{Q}\)) excitons. The corresponding real-space electron–hole probability densities are derived from the excitonic wavefunction. 
(c) Excitonic dispersion obtained from the solution of BSE, showing the five lowest branches along high-symmetry directions. The bright intravalley \textbf{K}–\textbf{K} exciton ($e_1$) occurs at \(\mathbf{Q}=0\) ($\boldsymbol{\Gamma}$), while finite-\(\mathbf{Q}\) states near \textbf{M} correspond to momentum-indirect excitons requiring phonon assistance for radiative recombination. 
(d–f) Real-space electron probability densities for the three lowest excitons ($e_1$, $e_2$, $e_3$), with the hole fixed on top of the carbon atom. The direct exciton ($e_1$) is compact and nearly isotropic, whereas the intervalley exciton ($e_3$) exhibits extended and anisotropic character consistent with their indirect nature.\\

\noindent \textbf{Figure 2.} Phonon mode resolved \textit{exc}-\textit{ph} couplings in 2D h-SiC. (a)-(f) Phonon-mode-resolved exciton-phonon coupling strength \( \left|G_{\beta\lambda,\nu}(\mathbf{Q},\mathbf{q})\right|^2 \) in meV$^2$ mapped across the 2D BZ of the phonon wavevector $\mathbf{q}$. For these plots, the initial exciton is fixed as the fundamental bright state ($\beta = e_1$, $\mathbf{Q}_{in} = \mathbf{\Gamma}$), and the coupling is summed over all available final exciton states ($\lambda$). The white hexagon denotes the first BZ boundary. To preserve the relative physical magnitudes between distinct phonon branches, all absolute coupling values are uniformly scaled by a factor of 1e4 rather than being independently normalized. Consequently, the varying limits on the color bars reflect the true differences in absolute scattering probabilities mediated by each mode.\\

\noindent \textbf{Figure 3.} Phonon mode resolved scattering rates and relaxation time in 2D h-SiC. Phonon-mode-resolved normalized scattering rates (in log) of excitons~1 and~5 at (a) 10~K and (b) 300~K, respectively. (c) Relaxation time (in log) for the lowest five excitons (1–5) at $\mathbf{Q}=0$ ($\mathbf{\Gamma}$) as a function of excitonic temperature $T_{\mathrm{exc}}$. Additionally, the $e_1$ valley exciton introduced in Fig. \ref{fgr:figure2}(a) physically corresponds to the primary bright state among the lowest $\mathbf{\Gamma}$-point eigenstates (1–5) analyzed here and in Fig. \ref{fgr:figure4}.\\

\noindent \textbf{Figure 4.} Excitonic linewidth variation with temperature. Excitonic temperature (T = T$_{\mathrm{exc}}$) dependence of the linewidths of excitons~1 and~5 at the $\mathbf{\Gamma}$ valley. The red dashed line represents the $\mathbf{\gamma}$ \textit{ab initio} model.\\

\noindent \textbf{Figure 5.} Indirect PL emission in 2D h-SiC. Phonon-assisted PL plotted on a logarithmic scale. The $x$-axes has been shifted by the energy of the bright exciton (2.97~eV). PL spectra at an excitonic temperature of 10 K. The inset shows the atomic vibrations (Si atoms in blue, C atoms in brown) are  of mode 5(A$^{\prime}$) and 6(A$^{\prime}$) by green and red arrows respectively.\\

\noindent \textbf{Figure 6.} Phonon-assisted PL emission from 2D h-SiC and h-BN.  Temperature-dependent normalized phonon-assisted PL spectra for (a) 2D h-SiC and (b) h-BN. The heatmaps display the PL intensity (logarithmic color scale) as a function of absolute emission energy (bottom axis) and detuning from the bright exciton $e_1$ (top axis) across a temperature range of $10$ to $300\,\mathrm{K}$. The spectra are normalized to their respective maximum emission peaks to emphasize the relative spectral weights of the excitonic and phonon-replica features. Dashed white lines trace the dominant emission channels, highlighting the direct emission ($e_1$) and the symmetry-allowed indirect emission ($e_2$), which is mediated by the $A^\prime$ phonon mode in 2D h-SiC and the LO-TO modes in h-BN. Moreover, the sideband peak in 2D h-SiC is $\sim$ 1.3 times larger than h-BN.

\clearpage
\onecolumngrid

\section*{Tables}
\addtocounter{table}{-1}
\begin{table}[ht]
\centering
\renewcommand{\arraystretch}{1.2}
\setlength{\tabcolsep}{8pt}
\begin{tabular}{|c|c|c|c|}
\hline
\textbf{Mode} & \textbf{Phonon irrep $\Gamma_{\mathrm{ph}}$} & \textbf{$\Gamma_{\mathrm{ph}}\otimes\Gamma_{\mathrm{light}}$} & \textbf{Transition} \\
\hline
1 & $A^{\prime\prime}$ & $A^{\prime\prime}$ & Forbidden\\
\hline
2 & $A^{\prime}$        & $A^{\prime}$ & Allowed\\
\hline
3 & $A^{\prime\prime}$        & $A^{\prime\prime}$ & Forbidden \\
\hline
4 & $A^{\prime}$        & $A^{\prime}$ & Allowed\\
\hline
5 & $A^{\prime}$ & $A^{\prime}$ & Allowed\\
\hline
6 & $A^{\prime}$        & $A^{\prime}$ & Allowed\\
\hline
\end{tabular}
\caption{\justifying Symmetry-allowed phonon-assisted optical transitions. Tensor-product decomposition between the in-plane light irrep, $\Gamma_{\mathrm{light}}=A^{\prime}$, and the phonon irreps at $\mathbf{K}$ for the six phonon branches considered. The last column represents the symmetry driven transitions.}
\label{tab:phonon}
\end{table}
\end{document}